# 5/4-approximation for Symmetric TSP

Alok Chauhan[*1], Madhusudan Verma[2]
[1]alok.chauhan@vit.ac.in, [2]madhusudan.verma2015@vit.ac
[1,2]VIT Chennai, India

**Abstract:** Travelling Salesman Problem (TSP) is one of the unsolved problems in computer science. TSP is NP-Hard. Till now the best approximation ratio found for symmetric TSP is 3/2 by Christofides' Algorithm more than thirty years ago. There are different approaches to solve this problem. These range from methods based on neural networks, genetic algorithm, swarm optimization, ant colony optimization etc. The bound is further reduced from 3/2 but for graphic TSP. A factor of 13/9 was found for Graphic TSP. A newly proposed heuristic called 2-RNN is considered here. It seems from experimental results that 5/4 is the approximation ratio. Upper bound analysis for approximation ratio is done for this heuristic and it confirms experimental bound of 5/4.

**Keywords:** Christofides' algorithm, 5-degree Minimum Spanning Tree, Approximation Ratio, Symmetric TSP, k-RNN.

## 1. Introduction

TSP asks to visit node exactly once and all vertices in a graph. TSP is APX-Hard. However, Held-Karp LP relaxation is conjectured to have bound of $4/3$. There is more general form of this problem known as Travelling Salesman Path Problem (TSPP) in which it is needed to find a path from two given points visiting all the vertices of graph exactly once. The best known algorithm for this problem is given by Hoogeveen . The bound found by this method is $5/3$. However it is conjectured to have an integrality gap of $3/2$ by the Held-Karp relaxation for this problem. One of the natural ways to attack this problem is to consider special cases of this problem. The most interesting is the Graphic TSP/TSPP. In Graphic TSP, we need to find a minimum cost circuit visiting vertices at least once. We can apply similar formulation to Graphic TSPP case. They are APX-hard, there are standard examples showing that the Held-Karp relaxation has a gap of at least $4/3$ in the TSP case and 3/2 in the TSPP case. A significant progress has been made in approximating the graphic TSP and TSPP in recent times. Oveis Gharan gave an approximation of $(3/2)\text{-}\varepsilon$ for Graphic TSP [3]. In which first an optimal solution of LP relaxation is computed. Then LP solution as $\lambda$-uniform distribution of spanning trees is written, followed by sampling of a Spanning Tree T from this distribution and at last a minimum cost matching on odd degree vertices of T is added. Following that, Mömke and Svensson obtained a significantly better approximation ratio of $\frac{14(\sqrt{2}-1)}{12\sqrt{2}-13} \approx 1.461$ for graphic TSP, as well as factor $3 - \sqrt{2} + \varepsilon \approx 1.586+\varepsilon$ for graphic TSPP, for any $\varepsilon > 0$. Above approach uses matching in a truly ingenious way. Instead of adding edges of a matching to a spanning tree to make it Eulerian, as it was done in previous approaches, the matching edges are added and removed. This process is guided by a so-called removable pairing of edges which essentially encodes the information on which edges can be simultaneously removed from the graph without disconnecting it. An approximation ratio of 5/4 for symmetric TSP is found in present work. This algorithm is simple to understand as well as easy to implement.

## 2. Motivation

The challenge to improve the approximation ratio obtained by Christofides is a big motivation. Since TSP has much wider applications, the need to work on this problem is felt.

## 3. k-RNN Algorithm

The algorithm is inspired by a new human centric co-existential philosophy propounded by Late Sri A Nagraj, India [10, 11]. Before explaining about 2-RNN, first let's understand its general form which is k-RNN [4]. The algorithm consists of the following steps:

Step 1: For every permutation of the k vertices $v_1, v_2, \ldots v_k$ create the partial tour T = $(v_1, v_2, \ldots v_k)$ and mark the vertices $v_1, v_2, \ldots v_k$ as visited.

Step 2: Set i = k. While there are unvisited vertices left: Select $v_{i+1}$ as the nearest unvisited neighbor of $v_i$ and append $v_{i+1}$ to T. If there are multiple nearest neighbors, select any. Mark $v_{i+1}$ as visited and increment i by 1.

Step 3: Among all n! / (n-k)! Tours found, select the shortest as the result.

2-RNN is k-RNN with k=2.

Now at first glance it seems similar to nearest neighbor algorithm, but the difference here is that instead of starting from a node, here we start from an edge.

Another observation about 2-RNN is that it can also be used to find the minimum of n*(n-1) Hamiltonian paths (open loop TSP).

## 4. Experimental Results for 2-RNN

| Dataset | Optimum | 1-RNN Result | 1-RNN Excess | 2-RNN Result | 2-RNN Excess |
|---|---|---|---|---|---|
| a280 | 2579 | 2975 | 15.35 | 2953 | 14.50 |
| berlin52 | 7542 | 8181 | 8.47 | 7968 | 5.65 |
| bier127 | 118282 | 133953 | 13.25 | 128589 | 8.71 |
| brazil58 | 25395 | 27384 | 7.83 | 27213 | 7.16 |
| brg180 | 1950 | 8890 | 355.90 | 2020 | 3.59 |
| ch130 | 6110 | 7129 | 16.68 | 6903 | 12.98 |
| ch150 | 6528 | 7113 | 8.96 | 7113 | 8.96 |
| d1291 | 50801 | 58681 | 15.51 | 58681 | 15.51 |
| d1655 | 62128 | 73369 | 18.09 | 72554 | 16.78 |
| d198 | 15780 | 17620 | 11.66 | 17405 | 10.30 |
| d493 | 35002 | 40186 | 14.81 | 40186 | 14.81 |
| d657 | 48912 | 60174 | 23.03 | 59310 | 21.26 |
| dantzig42 | 699 | 864 | 23.61 | 826 | 18.17 |
| eil101 | 629 | 746 | 18.60 | 743 | 18.12 |
| eil51 | 426 | 482 | 13.15 | 472 | 10.80 |
| eil76 | 538 | 608 | 13.01 | 598 | 11.15 |
| fl1400 | 20127 | 25115 | 24.78 | 24719 | 22.82 |
| fl417 | 11861 | 13887 | 17.08 | 13866 | 16.90 |
| fri26 | 937 | 965 | 2.99 | 959 | 2.35 |
| gil262 | 2378 | 2823 | 18.71 | 2767 | 16.36 |
| gr120 | 6942 | 8438 | 21.55 | 8335 | 20.07 |
| gr17 | 2085 | 2178 | 4.46 | 2178 | 4.46 |
| gr21 | 2707 | 3003 | 10.93 | 2958 | 9.27 |
| gr24 | 1272 | 1553 | 22.09 | 1400 | 10.06 |
| gr48 | 5046 | 5840 | 15.74 | 5561 | 10.21 |
| hk48 | 11461 | 12137 | 5.90 | 12031 | 4.97 |
| kroA100 | 21282 | 24698 | 16.05 | 24582 | 15.51 |
| kroA150 | 26524 | 31479 | 18.68 | 31320 | 18.08 |
| kroA200 | 29368 | 34543 | 17.62 | 34543 | 17.62 |
| kroB100 | 22141 | 25884 | 16.91 | 25255 | 14.06 |
| kroB150 | 26130 | 31611 | 20.98 | 31524 | 20.64 |
| kroB200 | 29437 | 35389 | 20.22 | 35283 | 19.86 |
| kroC100 | 20749 | 23660 | 14.03 | 23603 | 13.75 |
| kroD100 | 21294 | 24852 | 16.71 | 24603 | 15.54 |
| kroE100 | 22068 | 24782 | 12.30 | 24445 | 10.77 |
| lin105 | 14379 | 16935 | 17.78 | 16147 | 12.30 |
| lin318 | 42029 | 49201 | 17.06 | 49201 | 17.06 |
| linhp318 | 41345 | 49201 | 19.00 | 49201 | 19.00 |
| nrw1379 | 56638 | 68531 | 21.00 | 67873 | 19.84 |
| p654 | 34643 | 43027 | 24.20 | 42935 | 23.94 |
| pa561 | 2763 | 3279 | 18.68 | 3269 | 18.31 |
| pcb1173 | 56892 | 70115 | 23.24 | 69085 | 21.43 |
| pcb442 | 50778 | 58950 | 16.09 | 58682 | 15.57 |
| pr76 | 108159 | 130921 | 21.04 | 128749 | 19.04 |
| si1032 | 92650 | 94083 | 1.55 | 93981 | 1.44 |
| si175 | 21407 | 22000 | 2.77 | 21906 | 2.33 |
| si535 | 48450 | 50036 | 3.27 | 50032 | 3.27 |
| swiss42 | 1273 | 1437 | 12.88 | 1425 | 11.94 |

**Figure 1:** Results for 48 instances of the Symmetric TSP taken from TSPLIB [4].

## 5. Comparison with related work

| Algorithm | TSP Type | Approximation Ratio | Time Complexity |
|---|---|---|---|
| Christofides | Symmetric | $\frac{3}{2}$ | $O(n^3)$ |
| Truncated Generalized Beta distribution Based on Christofides' Algorithm | Symmetric | $\left(1+\frac{1}{2}\left(\frac{\alpha+1}{\alpha+2}\right)^{K-1}\right)$ | $O(n^4)$ |

| [9] | | where $\alpha >> 1$ is the shape parameter of TGB and K is the number of iterations | |
| --- | --- | --- | --- |
| 2-RNN [4] | Symmetric | $\frac{5}{4}$ | $O(n^4)$ |
| Random Sampling [3] | Graphic | $\frac{3}{2} - \in$ | unknown |
| Novel use of matching [5] | Graphic | $\frac{13}{9}$ | unknown |
| By ear-decomposition optimized using forest representations of hyper graphs[6] | Graphic | $\frac{7}{5}$ | Polynomial time |
| Finding a cycle cover with relatively few cycles for cubic bipartite graph [7] | Graphic | $\frac{9}{7}$ | Polynomial time |
| By consecutive path cover improvements [8] | Metric | $\frac{8}{7}$ | Polynomial time |

**Table 1:** Comparison of various TSP algorithms

# 6. Upper bound of 2-RNN approximation ratio for symmetric Euclidean space TSP

Given below is a proof for upper bound of 2-RNN approximation ratio which is found to be 1.25.

**Proof steps**

**Step 1**

**Lemma 1:** For any 4-node undirected complete graph G (V, E), V= {A, B, C, D} 2-RNN produces shortest Hamiltonian path of the G.

**Proof:** Let us assume that shortest path starts with edge AB then 2-RNN will choose smaller between BC and BD as next edge. Say, it is BC then resultant Hamiltonian path is ABCD. Let us assume BD is chosen instead of BC then resultant Hamiltonian path would be ABDC > ABCD. Hence it is proved.

**Lemma 2:** For any 5-node undirected complete graph G (V, E), V= {A, B, C, D, E} 2-RNN produces shortest Hamiltonian path of the G.

**Proof:** Let us assume that shortest path starts with edge AB then 2-RNN will choose smallest among BC, BD and BE as next edge. Say, it is BC then smaller between CD and CE is selected as next edge and say it is CD. Resultant Hamiltonian path is ABCDE. Other options:

  I. ABCED > ABCDE as CD < CE
 II. ABDCE > ABCDE as BC <BD, DE < CE as shown below:

    From triangle inequality:    BC+CE>=BE

                            BD+DE>=BE
                            hence CE > DE as BC < BD
 III. ABDEC > ABCDE as BD > BC, CE > CD
 IV. ABEDC > ABCDE as BE > BC
  V. ABECD > ABCDE as BE > BC, CE > DE

Hence it is proved.

**Step 2**

It is known that for every set of points in the plane, there exists a degree-5 MST [12]. Given a Euclidean minimum spanning tree T in which every vertex has degree at most 5, algorithm given below converts this tree to a tree in which every vertex has degree at most 4 [1]:

Let $V = \{v_1, \ldots, v_n\}$ be a set of $n$ points in the plane. Let $G$ be the complete graph induced by $V$, where the weight of an edge is the Euclidean distance between its endpoints.

TREE-4$(V, T)$ — *Find a degree 4 tree of $V$.*
1. Root the MST $T$ at a leaf vertex $r$.
2. **For** each vertex $v \in V$ **do**
3.     Compute the shortest path $P_v$ visiting $v$ and all its children.
4. Return $T_4$, the tree formed by the union of the paths $\{P_v\}$.

The tree T is rooted at an arbitrary leaf vertex. Since T is a degree-5 tree, once it is rooted at a leaf, each vertex has at most four children. For each vertex v, the minimum weight path $P_v$ visiting v and all of v's children (not necessarily starting at v) is computed. The final tree $T_4$ consists of the union of the paths $\{P_v\}$.

The cases when v has no children, one child, or two children are trivial. The case when v has 3 children can be thought of as equivalent to finding the shortest Hamiltonian path of a 4 node complete graph. Similarly, the case when v has 4 children can be thought of as equivalent to finding the shortest Hamiltonian path of a 5 node complete graph.

**Step 3**

We add following definition:

**Minimum tour spanning tree:** It is a 2-degree spanning tree obtained after removing any edge from the minimum tour generated by 2- RNN algorithm.

**Theorem 1:** Given a complete graph G, let TS be the minimum tour spanning tree and S is the 4-degree spanning tree derived from 5-degree minimum spanning tree of G by algorithm TREE-4(V,T), then

$$\text{Cost (TS)} < \text{Cost(S)} \dots\dots\dots\dots\dots\dots\dots(1)$$

**Proof:** TREE-4(V, T) considers a subset of vertices (either 1 or 2 or 3 or 4 or 5 number of vertices) of graph G. Imagine a complete sub graph consisting of this subset of vertices. Then, paths are generated covering all the vertices and shortest path among all is chosen. This process is repeated for all the subsets as per algorithm and finally the union of all such shortest paths is taken to generate a 4-degree spanning tree S of the graph G. Sub graphs of 1, 2 or 3 vertices are trivial cases and, interestingly, it has been shown that 2-RNN algorithm produces shortest Hamiltonian path for any 4 or 5 node complete graph (lemma 1, 2).

Now, we know that shortest Hamiltonian path problem (open loop TSP) is a non-decomposable problem. Algorithm TREE-4(V, T) can also be thought of as producing 4-degree spanning tree S due to union of local optimizations through 2-RNN for a non-decomposable problem, while algorithm 2-RNN also produces minimum tour spanning tree TS as global approximation for a non-decomposable problem.

Cost (TS) $\geq$ Cost(S) if and only if shortest Hamiltonian path problem for a complete graph is decomposable, hence

**Cost (TS) < Cost (S)**

**Step 4**

**Theorem 2:** Let T be the minimum tour generated by 2-RNN for graph G having n nodes and S be the 4-degree, ST of the given graph G derived from 5-degree MST of the graph G, by algorithm TREE-4(V, T) then

$$Cost(T) < \left(\frac{n}{n-1}\right) Cost(S) \dots\dots\dots\dots\dots\dots\dots\dots\dots\dots\dots (2)$$

**Proof**: Let TS be the minimum tour spanning tree generated by 2-RNN. There are n such trees, so summing n such trees,

$$\sum_{i=1}^{n} Cost(TS) < \sum_{i=1}^{n} Cost(S) \quad \text{(Apply theorem1)}$$

$$\Rightarrow (n-1) Cost(T) < n Cost(S)$$

$$\Rightarrow Cost(T) < \left(\frac{n}{n-1}\right) Cost(S)$$

**Step 5**

Following theorem is proved in [1]:

**Theorem 3:** Let MST be a minimum spanning tree of a set of points in $R^2$. Let S be the spanning tree output by TREE-4(V, T) algorithm, then

W (S) ≤ 1.25xW (MST)

Also, it is known that if TREE be the Minimum Spanning Tree of the given graph with n nodes and OPTIMAL be the length of optimal tour in graph visiting and starting at the same node [2] then,

Cost (TREE) ≤ (1- 1/n) OPTIMAL

**Theorem 4:** Let T be the minimum tour generated by 2-RNN for graph G and OPTIMAL is the length of optimal tour then

Cost (T) $< \dfrac{5}{4}$ OPTIMAL

**Proof:** By theorem 2,

$$Cost(T) < (n/(n-1)) Cost(S)$$

Since Cost (S) ≤ $\dfrac{5}{4}$ Cost(MST) by Theorem 3

Therefore $Cost(T) < 5/4(n/n-1)$ Cost (MST)

Since Cost (MST) ≤ $\left(1 - \dfrac{1}{n}\right)$ OPTIMAL

We get Cost (T) $< \frac{5}{4}\left(1-\frac{1}{n}\right)\left(\frac{n}{n-1}\right)$ OPTIMAL

$\Rightarrow$ **Cost(T)** $< \frac{5}{4}$ **OPTIMAL**

**Conjecture:** It is conjectured that the approximation ratio for k-RNN algorithm is $\frac{k^2+1}{k^2}$ for k>1.

## 7. Conclusion and Future Work

If we choose any two vertices for initial tour, the bound for the ratio between tour by 2-RNN and optimal is 5/4 . This can be baseline for finding the bound for the ratio if we choose k vertices as initial tour. Further research can be done to prove the above mentioned conjecture which may lead to a proof for P = NP. Though the time complexity of 2-RNN is O (n4) which is an order more than Christophides's algorithm O (n3) but at the same time 2-RNN is embarrassingly parallel in nature.